\documentclass[11pt,twoside]{article}
\usepackage{asp2006}
\usepackage{epsf}
\usepackage{graphicx}
\begin{document}

\title{Density Statistics of Compressible MHD Turbulence}

\author{A. Lazarian\altaffilmark{1}, G. Kowal\altaffilmark{1,2} \& A. Beresnyak\altaffilmark{1}}

\altaffiltext{1}{Astronomy Department, University of Wisconsin, Madison, WI 53706}
\altaffiltext{2}{Astronomical Observatory, Jagiellonian University, Krak\'ow, Poland}

\begin{abstract}
Density is the turbulence statistics that is most readily available from observations. Different regimes of turbulence correspond to different density spectra. For instance, the viscosity-damped regime of MHD turbulence relevant, for instance, to partially ionized gas, can be characterized by shallow and very anisotropic spectrum of density. This spectrum can result in substantial variations of the column densities. Addressing MHD turbulence in the regime when viscosity is not important over the inertial range, we demonstrate with numerical simulations that it is possible to reproduce both the observed Kolmogorov spectrum of density fluctuations observed in ionized gas by measuring scintillations and more shallow spectra that are obtained through the emission measurements. We show that in supersonic turbulence the high density peaks dominate shallow isotropic spectrum, while the small-scale underlying turbulence that fills most of the volume has the Kolmogorov spectrum and demonstrates scale-dependent anisotropy. The limitations of the spectrum in studying turbulence induce searches of alternative statistics. We demonstrate that a measure called "bispectrum" may be a promising tool. Unlike spectrum, the bispectrum preserves the information about wave phases.
\end{abstract}

\section{Density Fluctuations as a Probe of MHD Turbulence}

The paradigm of interstellar medium has undergone substantial changes recently. Instead of quiescent medium with hanging and slowly evolving clouds a turbulent picture emerged (see review by Ballesteros-Parredes et al. 2006, McKee \& Ostriker 2007 and ref. therein). With magnetic field being dynamically important and dominating the gas pressure in molecular clouds, this calls for studies of compressible magnetohydrodynamic (MHD) turbulence.

Key ideas in describing MHD turbulence can be traced back to Iroshnikov (1963) and Kraichnan (1965) work and the classical work that followed (see Montgometry \& Turner 1981, Higdon 1984, Montgomery, Brown \& Matthaeus 1987). A more recent model by Goldreich \& Sridhar (1995, henceforth GS95) has been successfully tested by numerical simulations (Cho \& Vishniac 2000, Maron \& Goldreich 2001, Cho, Lazarian \& Vishniac 2002, Cho \& Lazarian 2002, 2003, henceforth CL03).

Density statistics is the easiest to infer from observations. In comparison, velocity spectra require elaborate techniques to be used (see Lazarian 2006 and ref. therein). How informative can be density fluctuations for understanding of MHD turbulence is the subject of the present review. However, we should mention that for many important astrophysical applications, e.g. for interstellar chemistry, for star formation, for propagation of radiation etc. the density fluctuations themselves are crucially important.

Subsonic compressible MHD is rather well studied topic today. It is  suggestive that there may be an analogy between the subsonic MHD turbulence and its incompressible counterpart, namely, GS95 model. Therefore the correspondence between the the two revealed in CL03 is expected.

It could be easily seen, that in the low-beta case density is perturbed mainly due to the slow mode (CL03). Slow modes are sheared by Alfv\'en turbulence, therefore they exhibit Kolmogorov scaling and GS95 anisotropy for low Mach numbers. However, for high Mach numbers we expect shocks to develop. Density will be perturbed mainly by those shocks.

One can also approach the problem from the point of view of underlying hydrodynamic equations. It is well known that there is a multiplicative symmetry with respect to density in the ideal flow equations for an isothermal fluid (see e.g. Passot \& Vazquez-Semadeni, 1998). This assume that if there is some stochastic process disturbing the density it should be a multiplicative process with respect to density, rather than additive, and the distribution for density values should be lognormal, rather that normal. The aforementioned work shows that for 1D numerical simulations of high-Mach hydrodynamics the distribution is approximately lognormal, having power-law tails in case of $\gamma\neq 1$.

In MHD the above described symmetry is broken by the magnetic tension. However, numerical studies in Beresnyak, Lazarian \& Cho (2005, henceforth BLC05) show an approximate correspondence of the PDFs obtained with MHD simulations to the lognormal scaling.

With a high sonic Mach we expect a considerable amount of shocks arise. In a sub-Alfv\'enic case, however, we expect oblique shocks be disrupted by Alfv\'enic shearing, and, as most of the shocks are generated randomly by driving, almost all of them will be sheared to smaller shocks. The evolution of the weak shocks will be again governed by the sonic speed, and structures from shearing as in low Mach case should arise.

We also note that shearing will not affect probability density function (PDF) of the density, but have to affect its spectra and structure function (SF) scaling. In other words, we deal with two distinct physical processes, one of which, random multiplication or division of density in presence of shocks, affect PDF, while the other, Alfv\'enic shearing has to affect anisotropy and scaling of the structure function of the density. The numerical studies that we describe below confirm the theoretical considerations above.

\section{Viscosity-Damped MHD Turbulence as a Special Case}

Turbulence can be viewed as a cascade of energy from a large injection energy scale to dissipation at a smaller scale. The latter is being established by equating the rate of turbulent energy transfer to the rate of energy damping arising, for instance, from viscosity. Naively, one does not  expect to see any turbulent phenomena below such a scale.

Such reasoning may not be true in the presence of magnetic field, however. Consider magnetized fluid with viscosity $\nu$ much larger than magnetic diffusivity $\eta$, which is the case of a high magnetic Prantl number $Pr$ fluid. The partially ionized gas can serve as an example of such a fluid up to the scales of ion-neutral decoupling (see a more rigorous treatment in Lazarian, Vishniac \& Cho 2004, henceforth LVC04). Fully ionized plasma is a more controversial example. For instance, It is well known that for plasma the diffusivities are different along and perpendicular to magnetic field lines. Therefore, the plasma the Prandtl number is huge if we use the parallel diffusivity $\nu_{\|}\gg \nu_{\bot}$. A treatment of the fully ionized plasma as a high Prandtl number medium is advocated in Schechochihin et al. (2004, henceforth SCTMM).

The turbulent cascade in a fluid with isotropic $\eta$ proceeds up to a scale at which the cascading rate, which for the Kolmogorov turbulence, i.e. $v_l\sim l^{1/3}$, is determined by the eddy turnover rate $v_l/l$ gets equal to the damping rate $\eta/l^2$. Assuming that the energy is injected at the scale $L$ and the injection velocity is $V_L$, the damping scale $l_c$ is $L Re^{-3/4}$, where $Re$ is the Reynolds number $L V_L/\eta$.The corresponding scale $l_c$ varies from $\sim 10^{-2}$~pc for Warm Neutral Medium to $10^{-4}$ for molecular clouds (LVC04). It is evident, however, that magnetic fields at the scale $l_c$ at which hydrodynamic cascade would stop are still sheered by eddies at the larger scales. This should result in creating magnetic structures at scales $\ll l_c$ (LVC04). Note, that magnetic field in this regime is not a passive scalar, but important dynamically. Thus, GS95 turbulence range an inertial range in the partially ionized gas from the injection scale to $l_c$, but below $l_c$ is the domain of the viscosity-damped turbulence, where plasma/gas compressed within current sheets can contribute to the observed extreme scattering events and the formation of the Small Ionized or Neutral Structures (SINS) (Lazarian 2007). Most compressions arises in the situations when the magnetic fields exhibits perfect reversals.

Cho, Lazarian \& Vishniac (2002) reported a new viscosity-damped regime of turbulence using incompressible MHD simulations (see an example of more recent simulations in Fig.~1). In this regime, unlike hydro turbulence, motions, indeed, do not stop at the viscosity damping scale, but magnetic fluctuations protrude to smaller scales. Interestingly enough, these magnetic fluctuations drive small amplitude velocity fluctuations at scales $\ll l_c$.
CL03 confirmed these results with compressible simulations and speculated that these small scale magnetic fluctuations can compress ambient gas to produce ubiquitous tiny structures observed in ISM (see also Fig. 1b). According to the model in LVC04, while the spectrum of averaged over volume magnetic fluctuations scales as $E_B(k)\sim k^{-1}$, the pressure within intermittent magnetic structures increases with the decrease of the scale as $(\delta \hat{b})^2_k\sim k$, while the filling factor $\phi_k \sim k^{-1}$, the latter being consistent with numerical simulations.
\begin{figure}
  \centering
  \includegraphics[width=3.6cm]{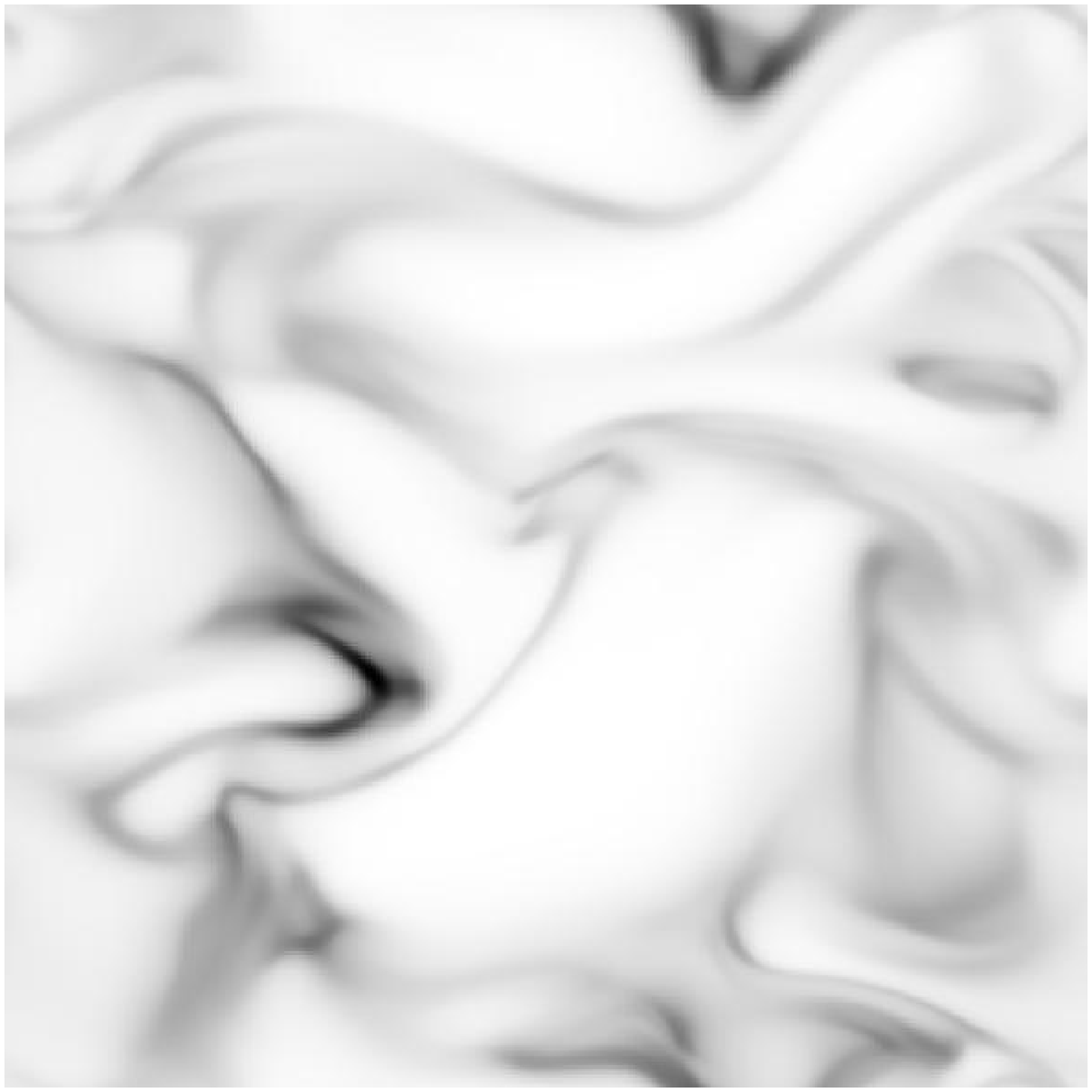}
  \includegraphics[width=0.67cm]{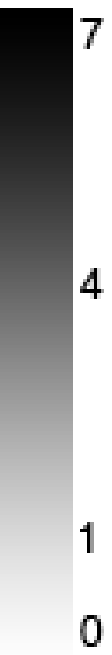}
  \includegraphics[width=5.0cm]{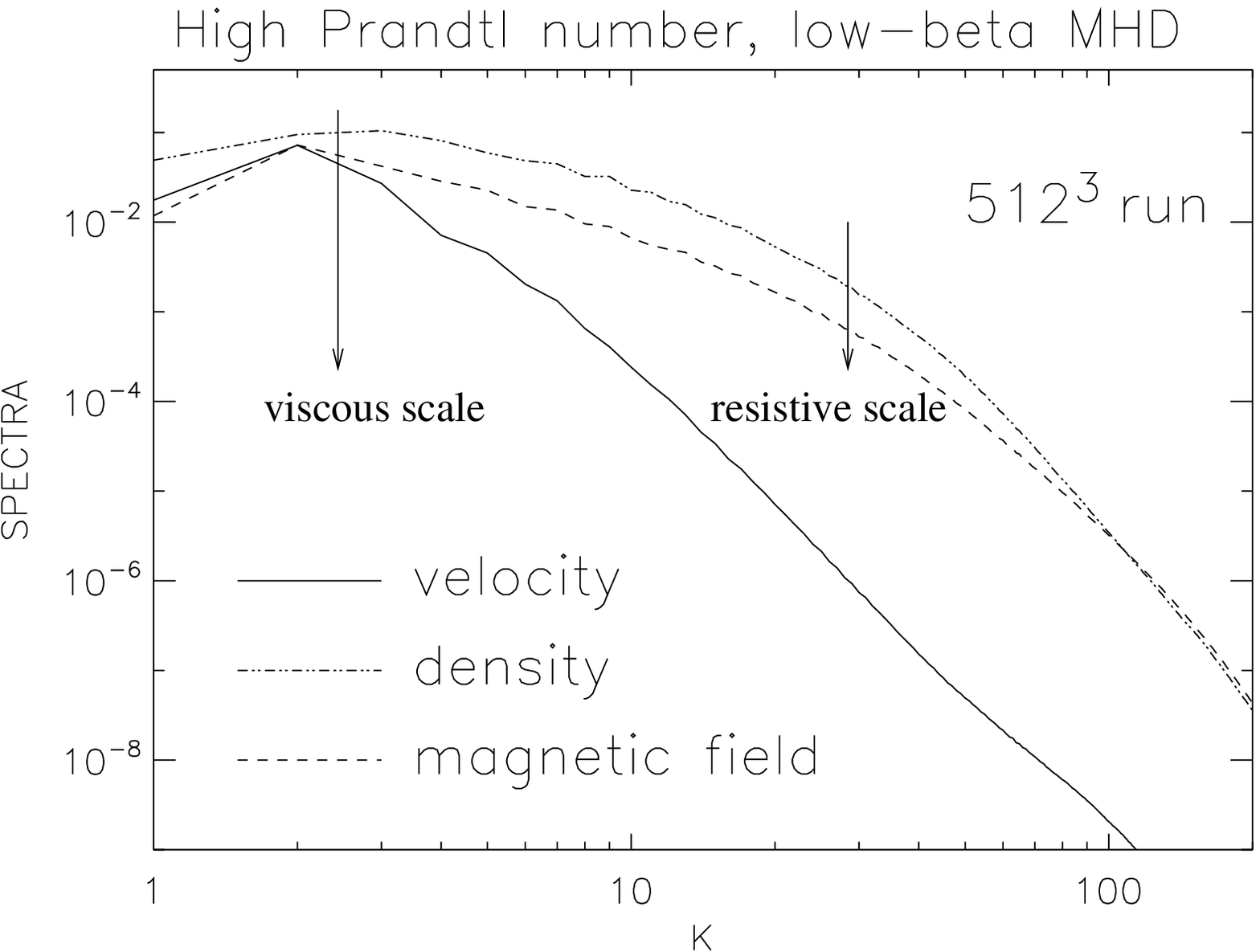}
\caption{ {\it Left}: Filaments of density created by magnetic compression in the slice of data cube of the viscosity-damped regime of MHD turbulence. {\it Right}: Spectra of density and magnetic field are similar, while velocity is damped. The resistive scale in this regime is not $L/Rm$ but $L {Rm}^{-1/2}$.(from Beresnyak \& Lazarian, in preparation).\label{visc}}
\end{figure}

In Fig.~\ref{visc} we show the results of our high resolution compressible MHD simulations that exhibit strong density fluctuations at the scales below the one at which hydrodynamic turbulence would be damped. According to  LVC04 model, the viscosity-damped regime is ubiquitous in turbulent partially ionized gas. Some of its discussed consequences, such as an intermittent resumption of the turbulence the fluid of ions as magnetic fluctuations reach the ion-neutral decoupling scale, are important for radio scintillations.

\section{Simulations 3D statistics and Simulated Observations}

\subsection{Spectrum}

The power spectrum (PS) of density fluctuations is an important property of a compressible flow. In some cases, the spectrum of density can be derived analytically. For nearly turbulent motions in the presence 
of a strong magnetic field, the spectrum of density scales similarly to the pressure, i.e. $E_{\rho}(k)\sim k^{-7/3}$ if we consider the polytropic equation of state $p=a\rho^{\gamma}$ \citep{biskamp03}. In weakly magnetized nearly
incompressible MHD turbulence, however, velocities convect density fluctuations passively inducing the spectrum $E_{\rho}(k)\sim k^{-5/3}$ \citep{montgomery87}. In supersonic flows, these relations are not valid anymore because of shocks accumulating matter into the local and highly dense structures. Due to the high contrast of density, the linear relation $\delta p = c_s^2 \delta \rho$ is no longer valid, and the spectrum of density cannot be related to pressure so straightforwardly. In addition, the strong asymmetry of density fluctuations suggests the need to analyze the logarithm of density instead of density itself.
\begin{figure}
 \plottwo{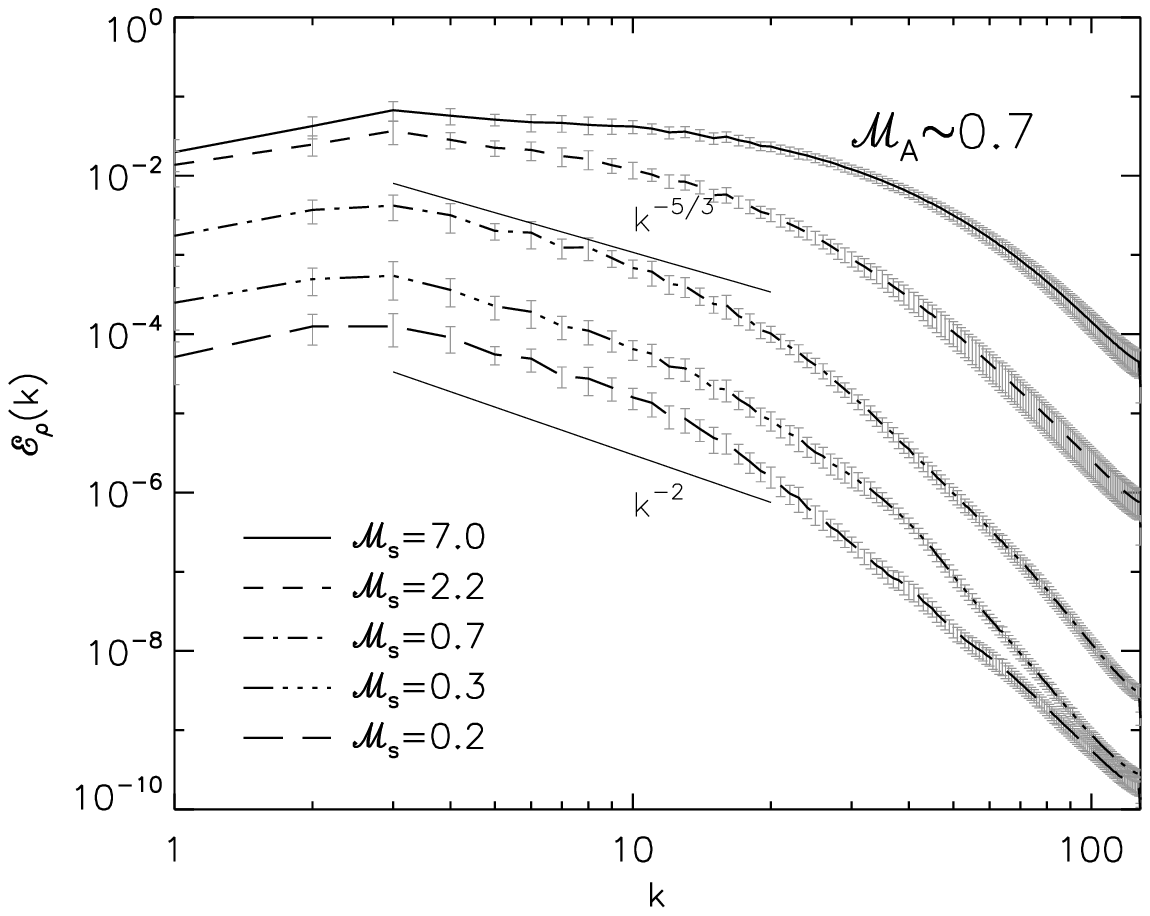}{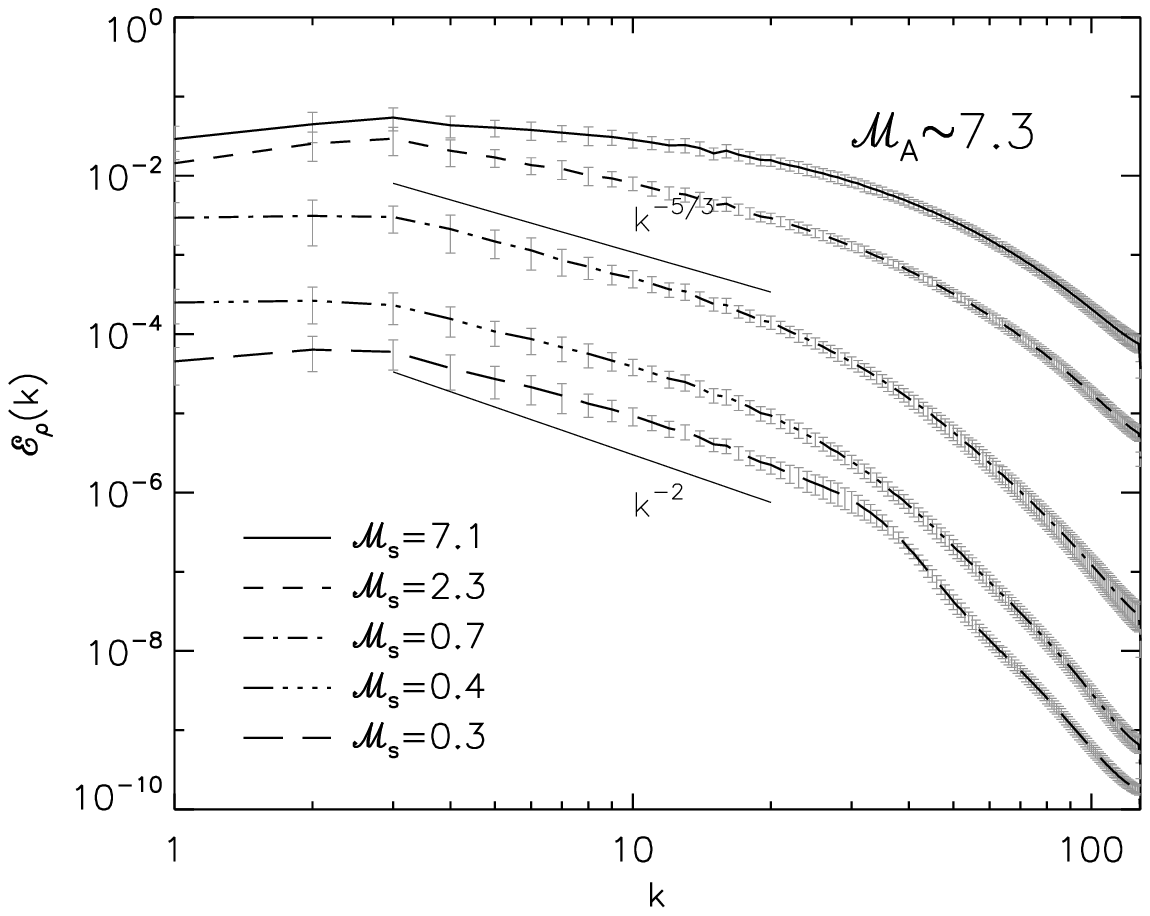}
 \caption{Spectra of density for experiments with different values of ${\cal M}_s$ and with ${\cal M}_A\sim0.7$ ({\em left panel}) and ${\cal M}_A\sim7.3$ ({\em right panel}) for models with medium resolution ($256^3$). Grey error bars signify the variance of spectra with time. The solid lines with slopes $-5/3$ and $-2$ cover the inertial range used to estimate the spectral indices of $\rho$ spectra \cite[from][]{kowal07}. \label{fig:spectra}}
\end{figure}

In Figure \ref{fig:spectra} we present the PS of fluctuations of density for models with different ${\cal M}_s$.  As expected, we note a strong growth of the amplitude of density fluctuations with the sonic Mach number at all scales. This behavior is observed both in sub-Alfv\'{e}nic as well as in super-Alfv\'{e}nic turbulence (see Fig. \ref{fig:spectra}). In Table \ref{tab:slopes} we calculate the spectral index of density $\alpha_\rho$ and the logarithm of density $\alpha_{\log \rho}$ within the inertial ranges estimated from the PS of velocity. The width of the inertial range is shown by the range of solid lines with slopes $-5/3$ and $-2$ in all spectra plots. It is estimated to be within $k \in (3, 20)$. In Table \ref{tab:slopes} we also show the errors of estimation which combine the error of the fitting of the spectral index at each time snapshot and the standard deviation of variance of $\alpha_{\rho, \log \rho}$ in time. The slopes of the density spectra do not change significantly with ${\cal M}_s$ for subsonic experiments and correspond to analytical estimations (about $-2.2$, which is slightly less than $-7/3$, for turbulence with ${\cal M}_A\sim0.7$ and about $-1.7$, which is slightly more than $-5/3$, for weakly magnetized turbulence with ${\cal M}_A\sim7.3$). Such an agreement confirms the validity of the theoretical approximations. Those, nevertheless, do not cover the entire parameter space. While the fluid motions become supersonic, they strongly influence the density structure, making the small-scale structures more pronounced, which implies flattening of the spectra of density fluctuations (see values for ${\cal M}_s>1.0$ in Table \ref{tab:slopes}) (see also BLC05). Although, the spectral indices are clearly different for sub- and super-Alfv\'{e}nic turbulence, their errors are relatively large (up to $\pm0.3$). These errors have been calculated by summing the maximum value of the uncertainty of fits for individual spectra at each time snapshot and the standard deviation of time variation of spectral indices. The uncertainty of fit contributes the most to the total error bar (about 60\%-70\%).
\begin{table}[t]
\caption{Slopes of the Power Spectrum of Density and the Logarithm of Density Fluctuations \cite[from][]{kowal07} \label{tab:slopes}}
\begin{center}
{\small
\begin{tabular}{ccc|ccc}
\noalign{\smallskip}
 {} & {${\cal M}_{A}\sim0.7$}  & {} & {} & {${\cal M}_{A}\sim7$} & {} \\
 {${\cal M}_{s}$} & {$\alpha_\rho$}  & {$\alpha_{\log\rho}$} & {${\cal M}_{s}$} & {$\alpha_\rho$}  & {$\alpha_{\log\rho}$}\\
\noalign{\smallskip}
\tableline
\noalign{\smallskip}
0.23$^{\pm0.01}$ & -2.3$^{\pm0.3}$ & -2.3$^{\pm0.3}$ & 0.26$^{\pm0.03}$ & -1.7$^{\pm0.3}$ & -1.7$^{\pm0.3}$\\
0.33$^{\pm0.01}$ & -2.2$^{\pm0.3}$ & -2.2$^{\pm0.3}$ & 0.36$^{\pm0.04}$ & -1.7$^{\pm0.3}$ & -1.7$^{\pm0.3}$\\
0.68$^{\pm0.03}$ & -2.0$^{\pm0.3}$ & -2.1$^{\pm0.3}$ & 0.74$^{\pm0.06}$ & -1.6$^{\pm0.2}$ & -1.6$^{\pm0.3}$\\
2.20$^{\pm0.03}$ & -1.3$^{\pm0.2}$ & -2.0$^{\pm0.2}$ & 2.34$^{\pm0.08}$ & -1.2$^{\pm0.2}$ & -1.6$^{\pm0.2}$\\
7.0$^{\pm0.3}$   & -0.5$^{\pm0.1}$ & -1.7$^{\pm0.2}$ & 7.1$^{\pm0.3}$   & -0.6$^{\pm0.2}$ & -1.5$^{\pm0.2}$\\
\noalign{\smallskip}
\tableline
\end{tabular}
}
\\
{\footnotesize
The values have been estimated within the inertial range for models with ${\cal M}_A\sim0.7$ ({\em left}) and ${\cal M}_A\sim7$ ({\em right}). Errors of spectral indices combine the errors of estimation at each time snapshot and the standard deviation of variance in time. Errors for sonic Mach numbers are the standard deviation of their variance in time calculated over the period starting from $t \ge 5$ to the last available snapshot.
}
\end{center}
\end{table}

\subsection{Anisotropies} 

\begin{figure}[t]
\centering
\plottwo{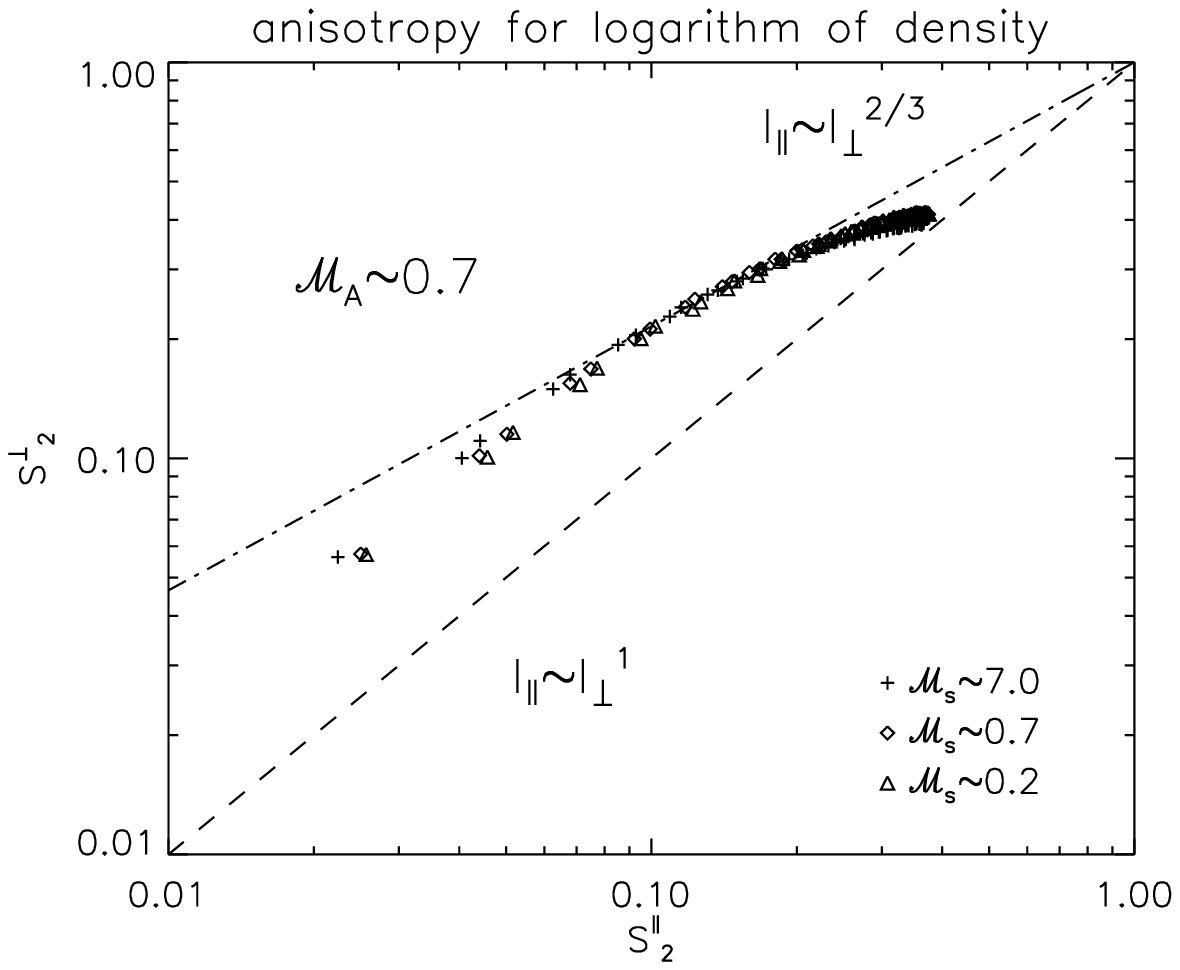}{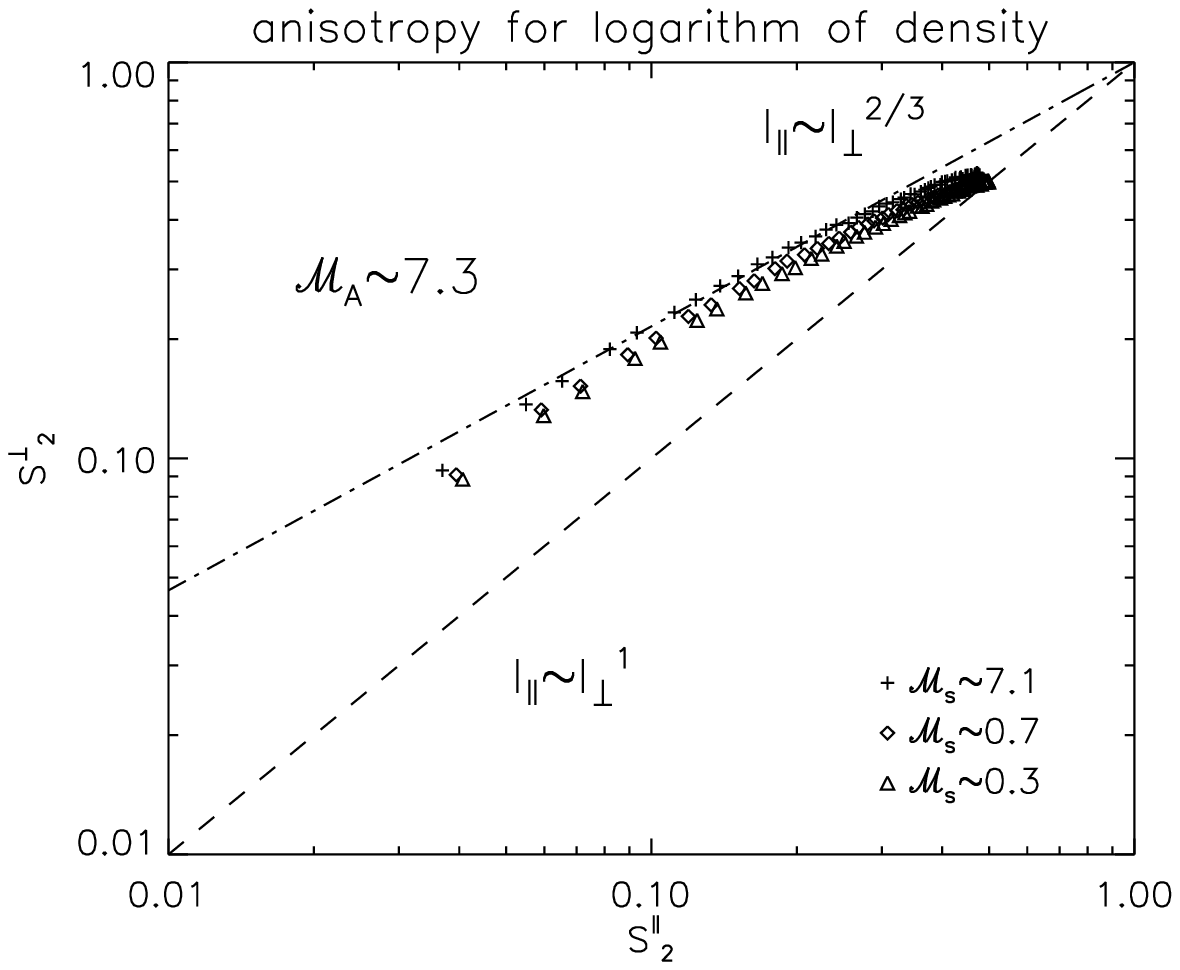}
 \caption{{\it Left}: Anisotropy for the 2$^{nd}$-order SF for the logarithm of
 density for ${\cal M}_A\sim7.3$; 
 {\it Right}: the same is for ${\cal M}_A\sim7.3$. The logarithm of density traces the anisotropies of turbulent magnetic field.}
\end{figure}
Another question is the anisotropy of density and the logarithm of density structures. For subsonic
turbulence it is natural to assume that the density anisotropies will mimic velocity anisotropies in GS95 picture. This
was confirmed in CL03, who, however, observed that for supersonic turbulence the contours of
density isocorrelation get round, corresponding to isotropy. BLC05, however, showed that anisotropies restore the
GS95 form if instead of density one studies the {\it logarithm of density}. This is due to the suppression of the influence
of the high density peaks, which arise from shocks. It is these peaks that mask the anisotropy of weaker, but more widely
spread density fluctuations.
 
In Figure \ref{fig:anisotropy} we show lines that mark the corresponding separation lengths for the second-order SFs parallel and perpendicular to the local mean magnetic field.\footnote{The local mean magnetic field was computed using the procedure of smoothing by a 3D Gaussian profile with the width equal to the separation length. Because the volume of smoothing grows with the separation length $l$, the direction of the local mean magnetic field might change with $l$ at an arbitrary point. This is an extension of the procedures employed in \cite{CLV}.} In the case of subAlfv\'{e}nic turbulence, the degree of anisotropy for density is very difficult to estimate due to the high dispersion of points. However, rough estimates suggest more isotropic density structures, because the points extend along the line $l_\parallel \sim l_\perp^1$. For models with ${\cal M}_A \sim 7.3$ the points in Figure~\ref{fig:anisotropy} have lower dispersion, and the anisotropy is more like the type from GS95, i.e. $l_\parallel \sim l_\perp^{2/3}$. In both Alfv\'{e}nic regimes, the anisotropy of density does not change significantly with ${\cal M}_s$. Plots for the logarithm of density show more smooth relations between parallel and perpendicular SFs. The dispersion of points is very small. Moreover, we note the change of anisotropy with the scale. Lower values of SFs correspond to lower values of the separation length (small-scale structures), so we might note that the logarithm of density structures are more isotropic than the GS95 model at small scales, but the anisotropy grows a bit larger than the GS95 prediction at larger scales. This difference is somewhat larger in the case of models with stronger external magnetic field (compare plots in the left and right columns of Figure \ref{fig:anisotropy}), which may signify their dependence on the strength of $B_\mathrm{ext}$. The anisotropy of $\log \rho$ structures is marginally dependent on the sonic Mach number, similar to the density structures. All these observations allow us to confirm the previous studies (see BLC05) that suggest that the anisotropy depends not only on the scale but on ${\cal M}_A$.

\subsection{Bispectrum}

Attempts to use multipoint statistics are a more traditional way to remove the constraints that the use of two point statistics, e.g. power spectra entails. Unfortunately, very high quality data is needed to obtain the multipoint statistics. Among multipoint statistics, bispectrum (see Scoccimarro 1997) seems the most promising. This is partially because it has been successfully used in the studies of the Large Scale Structure of the Universe.

Bispectrum is a Fourier transform of the three point correlation function and if the power spectrum $P({\bf k})$ is defined as
\begin{equation}
\langle \delta\rho({\bf k}_1)\delta
\rho({\bf k}_2)\rangle = P({\bf k})\delta_D ({\bf k}_1+{\bf k}_2)
\end{equation}
where $\delta_D$ is the Dirac delta function that is zero apart from the case when ${\bf k}_1+{\bf k}_2=0$, the bispectrum $B_{123}$ is
\begin{equation}
\langle \delta\rho({\bf k}_1)\delta \rho({\bf k}_2)\delta \rho({\bf k}_3)
\rangle=B_{123}\delta_D ({\bf k}_1+{\bf k}_2+{\bf k}_3).
\end{equation}

\begin{figure*}[t]
\centering
\plottwo{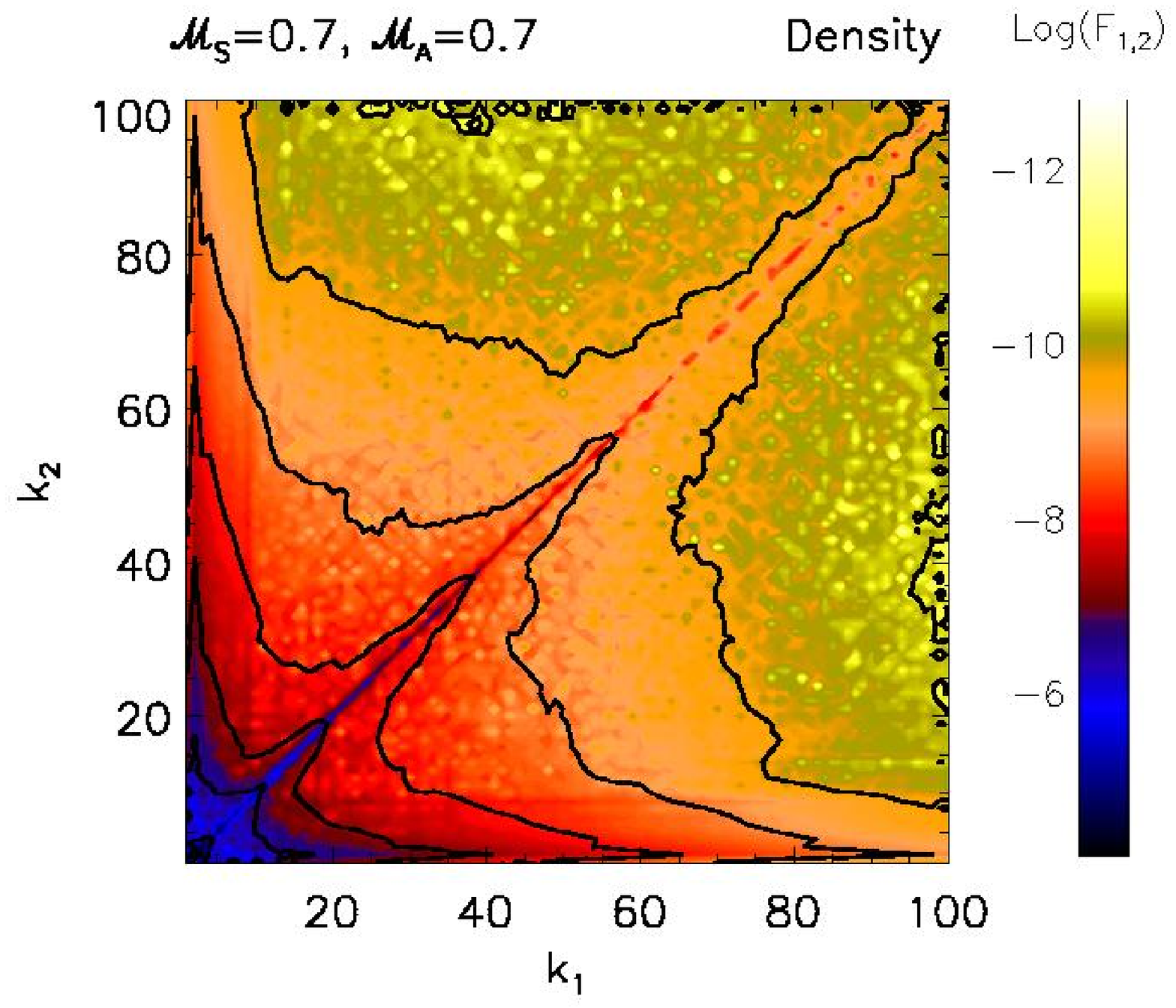}{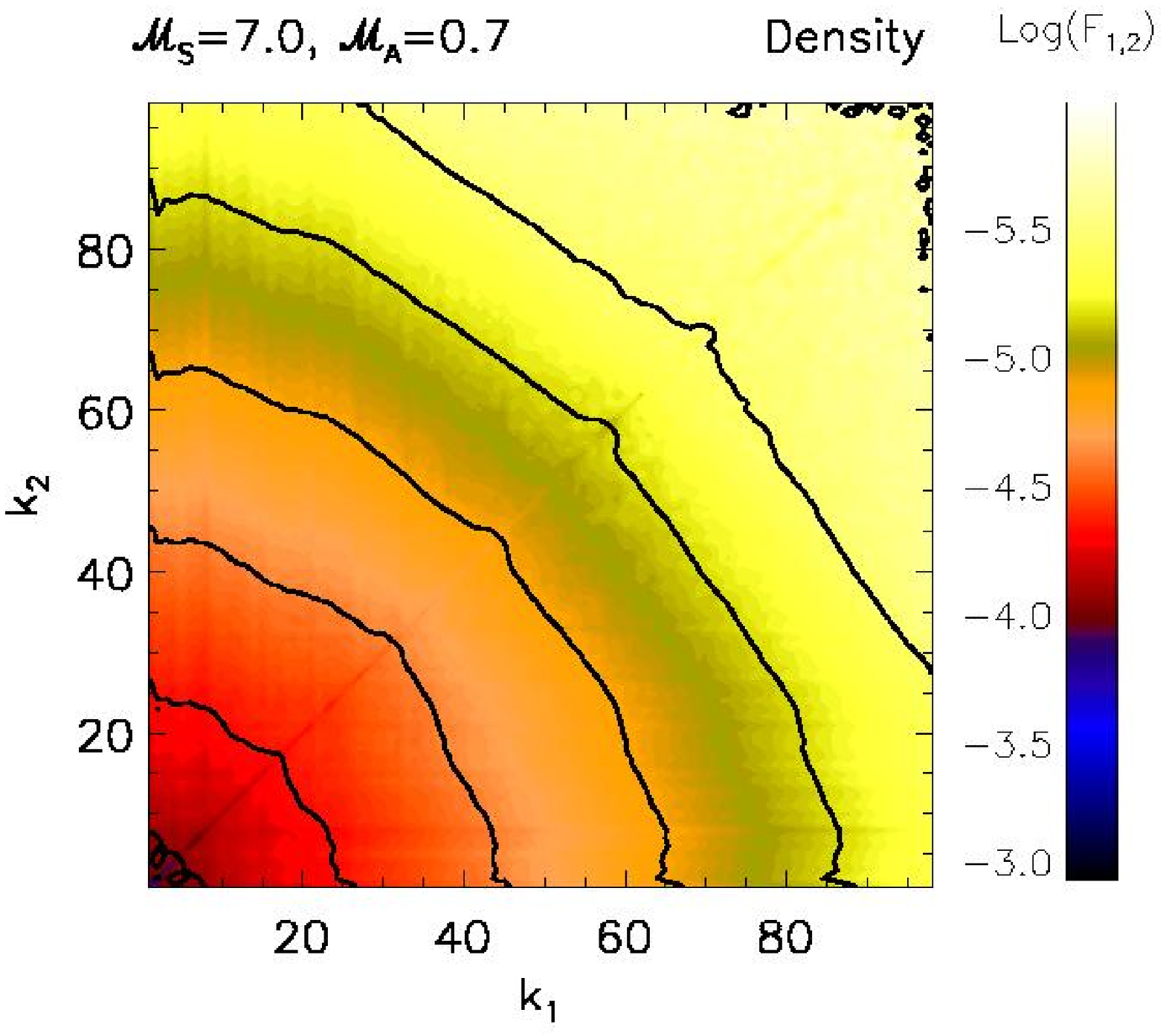}
\caption{The bispectra for density show the degree of correlation between $k_{1}$ and $k_{2}$. Here we compare different sonic numbers for strongly magnetized turbulence. The structure is clearly different for subsonic (left panel) and supersonic (right panel) cases. \label{fig:bispectra}}
\end{figure*}
The bispectra of density shown in Fig.~\ref{fig:bispectra} give information as to how shocks and magnetic fields effect turbulence. It has been shown by \cite[]{kowal07} and \cite[]{beresnyak05} that in supersonic turbulence shocks produce compressed density and a shallower spectrum. Looking at Fig.~\ref{fig:bispectra} it is clear that these shocks play a crucial role in the correlation of modes. The subsonic cases show little correlation between any points except the case of $k_{1}=k_{2}$. The compressed densities from supersonic turbulence are critical for correlations between frequencies since waves get closer together and therefore have a much higher interaction rate. We find an good agreement with \cite[]{kowal07} in that the density structures are affected by the presence of magnetic field. When there is a weak magnetic field present it is clear from Figure~\ref{fig:bispectra} that the system lacks the stronger correlations that are characteristic of the sub-Alf\'{v}enic models. For super-Alf\'{v}enic simulations correlations in frequency are not as readily made due to large dispersion of density structure. The bispectrum of supersonic hydro models are similar to the super-Alf\'{v}enic, supersonic cases. A 
measure related to bispectrum is {\it bicohence} (see Koronovskii \& Hramov 2002).  

\section{Summary}

Above we discussed the statistics of density in compressible MHD turbulence. We analyzed spectra, structure functions for experiments with different sonic ${\cal M}_s$ and Alfv\'{e}n ${\cal M}_A$ Mach numbers. Our results are as follows:

$\bullet$ The viscosity-damped regime of MHD turbulence relevant to partially ionized gas can be characterized by shallow and very anisotropic spectrum of density. This spectrum can result in large variations of the column densities.

$\bullet$ The amplitude of density fluctuations strongly depends on ${\cal M}_s$ both in weakly and strongly magnetized turbulent plasmas.

$\bullet$ The flattening observed in density spectra is due to the contribution of the highly dense small-scale structures generated in the supersonic turbulence.

$\bullet$ Fluctuations of the logarithm of density are much more regular than those of density. The logarithm of density exhibits the GS95 scalings and anisotropies.

$\bullet$ Bispectra show strong correlations for supersonic models than for subsonic ones. It suggests the importance of shocks in mode correlations.

{\footnotesize {\bf Acknowledgement} of the support of the NSF Center for Magnetic Self-Organization in Laboratory and Astrophysical Plasma.}


\begin{thebibliography}{}
\bibitem{} Ballesteros-Paredes, J.,  Klessen, R., Mac Low, M.
 \& Vasquez-Semadeni, E. 2006, in ``Protostars and Planets V'', University of Arizona Press, Tucson, p.63
\bibitem[Beresnyak et al. (2005)]{beresnyak05}
  Beresnyak, A., Lazarian, A. \& Cho, J., 2005, \apj, 624, L93, BLC05
\bibitem[Biskamp (2003)]{biskamp03}
  Biskamp, D., 2003, {\em Magnetohydrodynamic Turbulence}, Cambridge University Press
\bibitem{} Cho, J. \& Lazarian A. 2002, Phys. Rev. Lett., 88, number 24, 5001-(1-4)
\bibitem{} Cho, J. \& Lazarian, A. 2003, MNRAS, 345, 325-339
\bibitem[Cho, Lazarian \& Vishniac (2002)]{CLV}
  Cho, J., Lazarian, A. \& Vishniac, E.T. 2002, ApJ, 564, 291
\bibitem[Cho \& Vishniac (2000)]{cho00}
  Cho, J. \& Vishniac, E.~T., 2000, \apj, 539, 273
\bibitem[Elmegreen \& Scalo (2004)]{elmegreen04}
  Elmegreen, B. \& Scalo, J. 2004, \araa, 42, 211
\bibitem[Goldreich \& Sridhar (1995)]{GS95}
  Goldreich, P. \& Sridhar, H. 1995, ApJ {438}, 763
\bibitem{} Higdon, J.C. 1984, ApJ, 285, 109  
\bibitem{} Iroshnikov, P.S. 1963, AZh, 40, 742  
\bibitem[Kowal, Lazarian \& Beresnyak (2007)]{kowal07}
  Kowal, G., Lazarian, A. \& Beresnyak, A., 2007, \apj, 658, 423
\bibitem{} Koronovskii, A. \& Hramov, A. 2002, Plasma Phys. Rep. 28, 666
\bibitem{} Kraichnan, R. 1965, Phys. Fluids, 8, 1385
\bibitem{} Lazarian, A. 2007, in "SINS-Small Ionized and Neutral Strucutres in Diffuse ISM", eds. M. Haverkorn \& W.M. Goss, ASP 365, 324  
\bibitem{} Lazarian, A. 2006,  AIP 874, 301-315
\bibitem[Lazarian, Vishniac,\& Cho (2004)]{LVC}
  Lazarian, A., Vishniac, E., \& Cho, J. 2004, ApJ, 603, 180
\bibitem[Lazarian \& Beresnyak (2005)]{lazarian05}
  Lazarian, A. \& Beresnyak, A., 2005, in {\em The Magnetized Plasma in Galaxy Evolution}, Krak\'{o}w, Jagiellonian University Press, p. 56
\bibitem{} Maron, J. \& Goldreich, P. 2001, ApJ, 554, 1175 
\bibitem{}  McKee, C. \& Ostriker, E. 2007,  {\it ARA}\&{\it A},  45, 565
\bibitem{} Montgomrey D.C. \& Turner L. 1981, Phys. Fluids, 24, 825   
\bibitem[Montgomery et al. (1987)]{montgomery87}
  Montgomery, D., Brown, M.~R., Matthaeus, W.~H., 1987, JGR, 92, 282

\end{thebibliography}
\end{document}